\documentclass[final]{siamltex}
\usepackage{amsmath, amssymb, graphics,graphicx}
\usepackage{amscd}
\usepackage[dvips]{hyperref}
\usepackage{textcomp}

\newcommand{\threepartdef}[6]
{
	\left\{
		\begin{array}{lll}
			#1 & \mbox{if } #2 \\
			#3 & \mbox{if } #4 \\
			#5 & \mbox{if } #6
		\end{array}
	\right.
}

\title{Monte Carlo Portfolio Optimization for General Investor Risk-Return Objectives and Arbitrary Return Distributions: \break  a Solution for Long-only Portfolios}

\author{William T. Shaw\thanks{Department of Mathematics, King's College London, Strand, London WC2R 2LS ({\tt william.shaw@kcl.ac.uk}).}}

\begin{document}

\maketitle

\begin{abstract}
We develop the idea of using Monte Carlo sampling of random portfolios to solve portfolio investment problems. In this first paper we explore the need for more general optimization tools, and consider the means by which constrained random portfolios may be generated. A practical scheme for the long-only fully-invested problem is developed and tested for the classic QP application. The advantage of Monte Carlo methods is that they may be extended to risk functions that are more complicated functions of the return distribution, and that the underlying return distribution may be computed without the classical Gaussian limitations. The optimization of quadratic risk-return functions, VaR, CVaR, may be handled in a similar manner to variability ratios such as Sortino and Omega,  or mathematical constructions such as expected utility and its behavioural finance extensions. Robustification is also possible. Grid computing technology is an excellent platform for the development of such computations due to the intrinsically parallel nature of the computation, coupled to the requirement to transmit only small packets of data over the grid. We give some examples deploying GridMathematica, in which various investor risk preferences are optimized with differing multivariate distributions. Good comparisons with established results in Mean-Variance and CVaR optimization are obtained when ``edge-vertex-biased'' sampling methods are employed to create random portfolios. We also give an application to Omega optimization. 
\end{abstract}



\pagestyle{myheadings}
\thispagestyle{plain}
\markboth{W.T.SHAW}{Monte Carlo Portfolio Optimization for General Objectives and Distributions}

\begin{center}
{\it You can't be serious, man. You can not be serious!}
John McEnroe, 1981.
\end{center}

\section{Introduction}
The management of financial portfolios by the methods of quantitative risk management  requires the solution of a complicated optimization problem, in which one decides how much of a given investment to allocate to each of $N$ assets - this is the problem of assigning weights. There are many different forms of the problem, depending on how large $N$ is ($2-2000+$!), whether short-selling is allowed, investor preferences, and so on.  In the simplest form of the problem, which is still relevant to traditional fund management, the weights are non-negative (you cannot go ``short'') and are normalized to add to unity. So if we have an amount of cash $P$ to invest and $N$ assets to choose from, the weights are $w_i, i=1,\dots,N$ satisfying 
\begin{equation}
w_i \geq 0
\end{equation}
\begin{equation}
\sum_{i=1}^N w_i = 1
\end{equation}
This {\it long-only} form of the problem is of considerable importance, but not the only one of interest. Additional constraints are often specified, and this will be allowed for in the following discussion. The development of optimized Monte-Carlo tools with the addition of short-selling is under investigation. 

The simplest and very ancient solution to this problem is to take $w_i = 1/N$, which is equal allocation, and has been understood for hundreds of years as a principle. The amount of cash invested in asset i is then, in general, $P w_i$ and in the equal-allocation strategy it is $P/N$. The oldest currently known statement of this principle is due to one Rabbi Isaac bar Aha, who in an Aramaic text from the 4th Century, stated his equal allocation strategy as: {\it A third in land, A third in merchandise, A third in cash  (``a third at hand'')}. The usefulness of this robust strategy continues to be explored - see for example the beautiful study by DeMiguel {\it et al} \cite{oneovern} and references therein.

It has become an in-joke to state that the underlying mathematical theory did not move on very quickly after the Rabbi. There {\it were} sporadic references in literature. Some proverbial mis-translation of {\it Don Quixote} is sometimes given as: {\it Tis the part of wise man to keep himself today for tomorrow, and not venture all his eggs in one basket}, but there is in fact no evidence in the original text of eggs. Shakespeare illustrated the concept of diversification around the same time in {\it The Merchant of Venice}, when Antonio told his friends he wasn't spending sleepless nights worrying over his commodity investment: {\it Believe me, no. I thank my fortune for it, My ventures are not in one bottom trusted, Nor to one place; nor is my whole estate upon the fortune of this present year. Therefore my merchandise makes me not sad}. It helps to know that ``bottom'' refers to a ship's hull, and that the diversification here is intended to prevent the entire cargo going down with a single ship. But note that Shakespeare had quietly added temporal diversification. 

There is also a contrarian thread of thought\footnote{My thanks to Michael Mainelli for making me aware of this.}: some people follow Mark Twain's contrarian, anti-portfolio investment strategy (from {\it Pudd'nhead Wilson}) and seek to concentrate on the next big thing: ``Behold, the fool saith, `Put not all thine eggs in the one basket' - which is but a manner of saying, `Scatter your money and your attention', but the wise man saith, `Put all your eggs in the one basket and - watch that basket'.'' A similar quote is also attributed to Andrew Carnegie, and it is often claimed that Warren Buffett leans towards a similar philosophy, based on several observations and his infamous quote:  {\it Concentrate your investments.  If you have a harem of 40 women you never get to know any of them very well.}
In the aftermath of the financial crisis, the following anti-diversification theme seems even more pertinent\footnote{I apologise for not being able to quote the source of this idea - seen once on the web and not found again - correction requested.}: {\it If you are stuck on a desert island and need a drink of water, and know that one of the three streams is poisoned, but not which one, would you take a third of your need from each stream?} Such debates will continue.  
 
Putting aide the cultural discussion, the mathematical theory moved on in the post-WWII work of Markowitz and collaborators, (see for example \cite{markow1,markow2,markow3}). My own understanding is that Markowitz was in fact interested in the control of downside risk (the semivariance), but given the technology then available he used variance as a proxy for risk. where the idea was formulated as (in the simplest version), the minimization of risk subject to a return goal. This resulted in a problem to minimize functions of the form
\begin{equation}
\sum_{i=1}^N\sum_{j=1}^N w_i w_j C_{ij} - \lambda\sum_{i=1}^N R_i w_i
\end{equation}
where $C_{ij}$is the covariance matrix of the assets and $R_i$ are the expected returns. With the summation convention due to Einstein, this is usually written in either of the more compact forms
\begin{equation}
w_i w_j C_{ij} - \lambda R_i w_i = w^T.C.w-\lambda R.w \label{qpeqn}
\end{equation}
and the minimization is subject to the constraints of equations (1,2) and indeed any other additional conditions (e.g. sector exposure, upper and lower bounds etc.)
The quantity $\lambda$ is a Lagrange multiplier expressing an investor's risk-return preference, and there are other ways of writing down the problem.

This form of the problem, known as constrained Quadratic Programming (QP), is highly restrictive. First there is the underlying idea that the risk is characterized solely by the variance. Implicit in this is the idea that the distribution of returns is essentially Gaussian. Second, if one thinks more generally in terms of maximizing expected Utility, Eqn.~(\ref{qpeqn}) only emerges if one either approximates, or considers exponential utility and again Gaussian returns. Third, the answers are extremely unstable with respect to changes in covariance or expected return. Fourth, the dependency model between returns is quite explicitly Gaussian, expressed by the covariance. Pathological covariances with zero or near-zero eigenvalues cause an instability, and optimal configurations may have zero weights in some assets and not in others. It will be important to realize that interior-point solutions are quite rare in practical examples, and optima frequently sit on boundary planes and vertices. 

There are other complications - investors might need more complex constraint sets and/or allow for short-selling, where some weights are negative, especially for hedge fund applications. These are fairly easily admitted into classical computation techniques, but relaxing all the other restrictions is problematic. Recent work on Robust Optimization allows for some uncertainty in the parameters, for example, with a QP framework. See for example Goldfarb et al \cite{goldfarbrob} and references therein, \cite{zuev} for minimal eigenvalue management. 

Investor preferences might not be expressed in terms of variance-return structures. Expected Utility is a common theme in more mathematical studies, but there are also specific investment functions that are written down without explicit reference to utility. Value at Risk and coherent variations are often considered, but remain controversial. The measures of VaR and CVaR have attracted a lot of attention following the pioneering work of Rockafellar and Uryasev  \cite{cvaropt}. This work, which I shall refer to collectively as the RU-method, opened up a whole new area of research into the optimization of full distributional measures of risk (i.e. not just working with simple moments). From our point of view a key feature is the modelling of the return distribution by Monte Carlo sampling of returns. This should be distinguished from what is being proposed now, in that we wish to take the {\it additional} step of {\it simulating random portfolios}. The RU-method is ingenious in that it maps the VaR, CVAR optimization problem into an LP problem solvable by well-established methods. Nevertheless it has the awkward feature that the mapping generally involves the creation of one constraint for each Monte Carlo realization of the portfolio, and therefore raises issues of computational scaling as one tries to model the return distribution more accurately, and especially when treating low-quantile tail risk. An alternative avoiding this issue has been proposed relatively recently by Iyengar and Ma \cite{cvarnew}.

But the choice of investor objective functions does not stop with VaR and CVaR. We will remark on other parts of the zoo\footnote{The intention here is not to be comprehensive, but I apologise for leaving out functions felt by others to be important. I just want the reader to appreciate that there is a very large zoo indeed, to lure you into the idea that it would be helpful to have a single method that could do them all.}. Investment performance has been measured by a variety of financially-motivated ideas. Expressions based on the Sharpe ratio abound. A mathematically neat characterization of one family has been given by Farinelli and Tibiletti \cite{variability}, who define a one-sided performance ratio as a function of a benchmark $b$ as follows:
\begin{equation}
\Phi(b,p,q) = \frac{E[((X-b)^+)^p]^{1/p}}{E[((b-X)^+)^q]^{1/q}}
\end{equation}
The choice $(p,q) = (1,2)$ recovers the Sortino ratio and $(p,q) = (1,1)$ is the ``Omega'' function \cite {omega}, discussed in more detail later. Mathematicians on the other hand often prefer to work with expected utility based on a choice of utility function, and the behavioural finance idea can be represented through prospect functions \cite{behaviour}, which this author regards as expressible as two-sided utility.

All these approaches are functions of the full probability distribution in general and generally speaking one has to work hard on a case by case basis to do the optimization. Identification of a gradient function to unleash classical methods may be difficult, so a universal gradient-free approach may be very useful. 

\subsection{Summary}
In general we have some mapping from the distribution of portfolio returns to the real numbers that is a function of the weighs of the assets in the portfolio.  Our goal may be stated as follows. For general choices of:
\begin{enumerate}
\item  asset-price evolution model or price probability distribution;
\item  inter-asset dependency model;
\item portfolio constraints;
\item investor objective function (including Quadratic Risk-Return, Pure down-sided risk measures, ``Omega'', ``Sortino ratio'', Utility, Behavioural Finance Prospect...);
\end{enumerate}
with the possibility of requiring robust results in the light of data uncertainty,  to evaluate the optimal weights in a portfolio.

The proposed solution\footnote{It is at this point the reader may refer back to my quote from McEnroe.} is as follows. The method is to find the optimum portfolio by a simple but extensive computational search amongst randomly generated portfolios, using a pre-created collection of portfolio return samples (as per the RU-method), but then 
\begin{itemize}
\item simulation of the possible weights in a careful manner;
\item direct calculation of the risk function for each weight;
\item selection of the weights by taking the best outcome;
\item exploiting parallel-processing technology to create a feasible system for large problems, by a trivial {\it best of the best} model, where different threads are fed the same return distribution but different seeds for the random portfolio generation. 
\end{itemize}
The method is summarized as Monte Carlo Portfolio Optimization in a Parallel Environment (MC-POPE). 

We wish to emphasize that the use of Monte Carlo methods for creating portfolio return realizations in connection with optimization is not new - this part of the concept was certainly explicit in the work of Rockafellar and Uryasev \cite{cvaropt} on CVaR optimization. Also, the generation of {\it random portfolios} for assisting portfolio analysis in the context of diverse types of risk analysis is now available as a commercial tool - see the website for ``PortfolioProbe'' for example\cite{probe}. However, we believe that the combination of random portfolio sampling linked to risk optimization, with the means of random portfolio generation itself optimized for the risk optimization problem, is an approach rather in its infancy, and this paper is a contribution to that thread of thought.

The parallel environments employed for our basic research are:

\begin{itemize}
\item  Grid Mathematica for idea development, testing, proof-of-concept and small-scale deployment.
\item GPU cards running CUDA or OpenCL for larger-scale problems.
\end{itemize}

It will be evident that the plan involves existing and well-known mathematics apart from the notion of efficiently sampling random portfolios, and it is to this that we now turn, and devote the bulk of our analysis. 

\section{The nature of optimal portfolios}
Monte Carlo simulation has in the past largely focused on the calculation of an expectation, and the simulation is employed to generate a numerical realization of a measure. Here we are trying to locate optima so the importance of the measure changes somewhat and we have some latitude about what to do. But there are bad measures and good measures and the meaning of this is a matter linked to the particular nature of non-linear optimization. To get a grip on this we will introduce a simple test problem in QP form. Consider the case of zero expected return and the pure variance-minimization problem  with $N=3$ defined by the parametrized covariance given by
\begin{equation}
C_{ij} = \left(
\begin{array}{ccc}
 64 & 120 r & 25 \\
 120 r & 225 & 50 \\
 25 & 50 & 100
\end{array}
\right)
\label{cova}
\end{equation}
where $-1 \leq r \leq 1$.
The solution to this problem as a function of the correlation parameter $r$ can be found analytically (given later) or with any number of off-the-shelf QP solvers and the results look like this, where we plot the 3 weights as a function of $r$:

\begin{figure}[hbt]
\begin{center}
\includegraphics[scale=1]{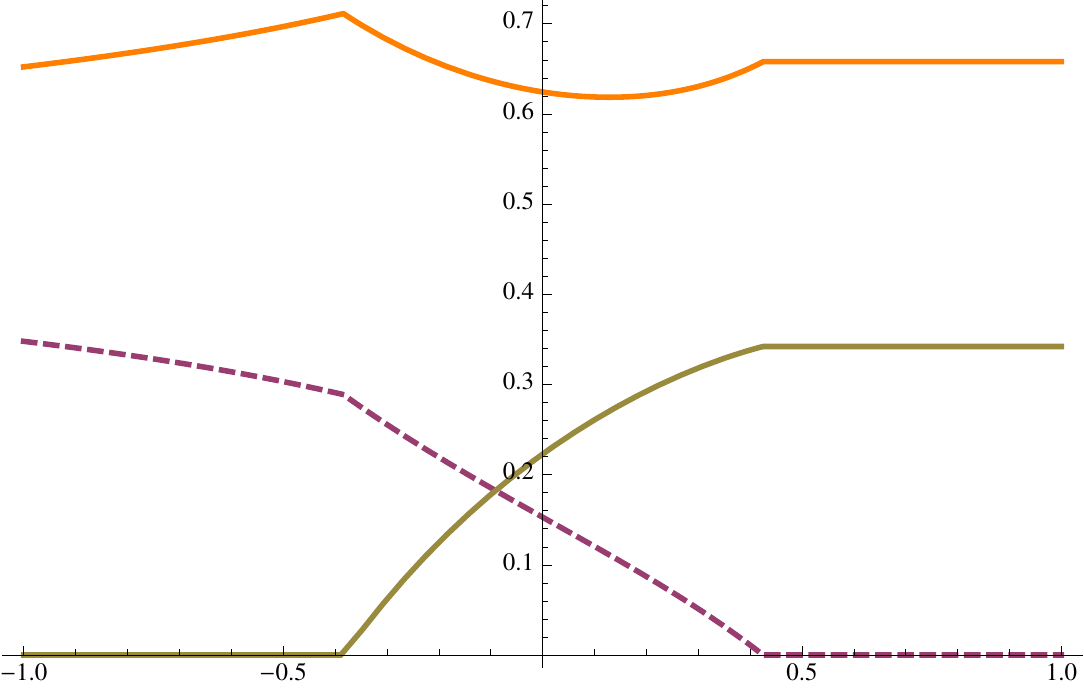}
\end{center}
\caption{Minimum variance weights}\label{weights}
\end{figure}
The details of this example do not matter yet. What matters is that for some $r$ the solution is an interior point where all weights are non-zero and when $r$ is close to $\pm 1$ one of the weights is zero. This is a typical feature of QP solutions. While LP problems always have solutions on the boundary of the feasible region, but QP problems may be interior point but will frequently have solutions with several zeroes in the weights for particular assets. We will return to this presently. 

\section{Random portfolios}
We will start with a basic approach, based on the comments at www.excel-modeling.com\footnote{The website as of August 2010 reports this approach.}, and while it is not what is ultimately needed, it provides a useful clue to improve matters. It should be noted that no criticism of the general integrity of the models on that web site is intended, and the author recognizes that their public web site might just contain a simple example for illustration only. I quote from the web site of excel-modeling.com: {\it 1. Generate 3 random numbers. 
2. Assign the ith random number divided by the sum of the three random numbers as the weight for stock i. The procedure above ensure each weight will be fairly distributed.}

This is an interesting idea. However, the statement at the end is in fact false, {\it if} the underlying random numbers are uniform on $(0,1)$\footnote{The company quoted may well have more sophisticated algorithms or different underlying distributions} - we shall see later what distributions make this procedure more sensible. Such a  procedure creates an uneven, biased distribution. It is easiest to see what goes wrong with a picture that shows the biased distribution of 4,000 samples of the simplex:
\begin{equation}
w_1 + w_2 + w_3 = 1\ , w_i \geq 0
\end{equation}
\begin{figure}[hbt]
\begin{center}
\includegraphics[scale=0.6]{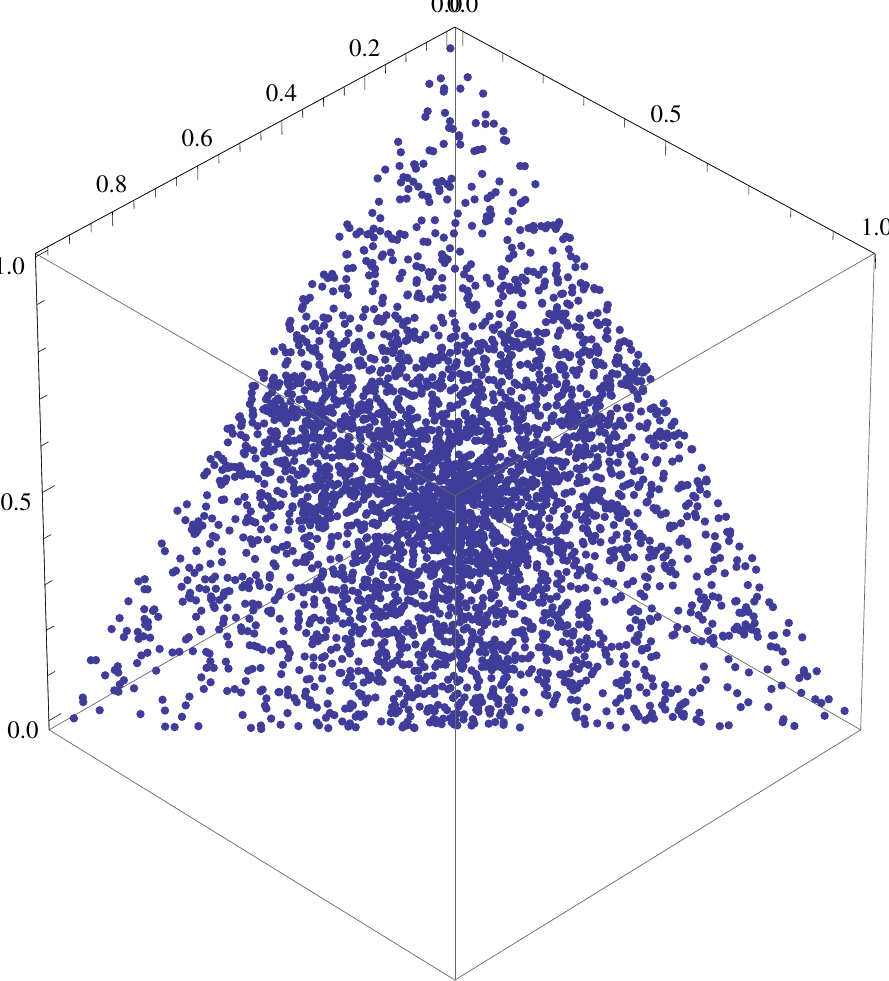}
\end{center}
\caption{Simple biased simplicial sample}\label{biased}
\end{figure}
The bias is evident from the scatter plot, that reveals that the distribution has a peak at the equally-weighted location, and subsidiary peaks in the middle of the edges. There is strong bias away from the vertices. This gets worse as you increase the dimension, as can be checked by some simple numerical experimentation. 

We will be unlikely to find correct optima buried on any vertices or edges using this method, which should be avoided in the view of this author.  Note, however, that we wish to reserve the option to bias, but in a different way. I will return to this later.

The next task is to properly characterize the methods for generating un-biased random portfolios with the appropriate constraints. I experimented with this for a while and came up with two methods, only to find the basic ideas were all properly written down over two decades ago. The theory for the simplest case was all written down by Luc Devroye in Chapter 5 of his beautiful book {\it Non-uniform Random Variate Generation}\cite{luc}. The idea of so-called ``simplicial sampling'' is also at the heart of many copula sampling methodologies. I will explain the three equivalent methods directly, but the reader is urged to consult the Devroye text. 

\subsection{Even simplicial sampling method 1}
This is based on a simple gap distribution. To get $k$ weights satisfying the constraints, generate $k-1$ points in the unit interval, sort them in increasing order and then consider the gaps, including the end points at zero and one. Note that this requires a sort in order to define the gaps from left to right sequentially. In the case $N=3$ we sample two points $U_1, U_2$, let $V_1 = {\rm min}(U_1, U_2), V_2 = {\rm max}(U_1, U_2)$ and just write down
\begin{equation}
(w_1, w_2, w_3) = (V_1, V_2-V_1, 1-V_2) \ .
\end{equation}
The result is then an beautifully even distribution (shown with 2000 points):
\begin{figure}[hbt]
\begin{center}
\includegraphics[scale=0.6]{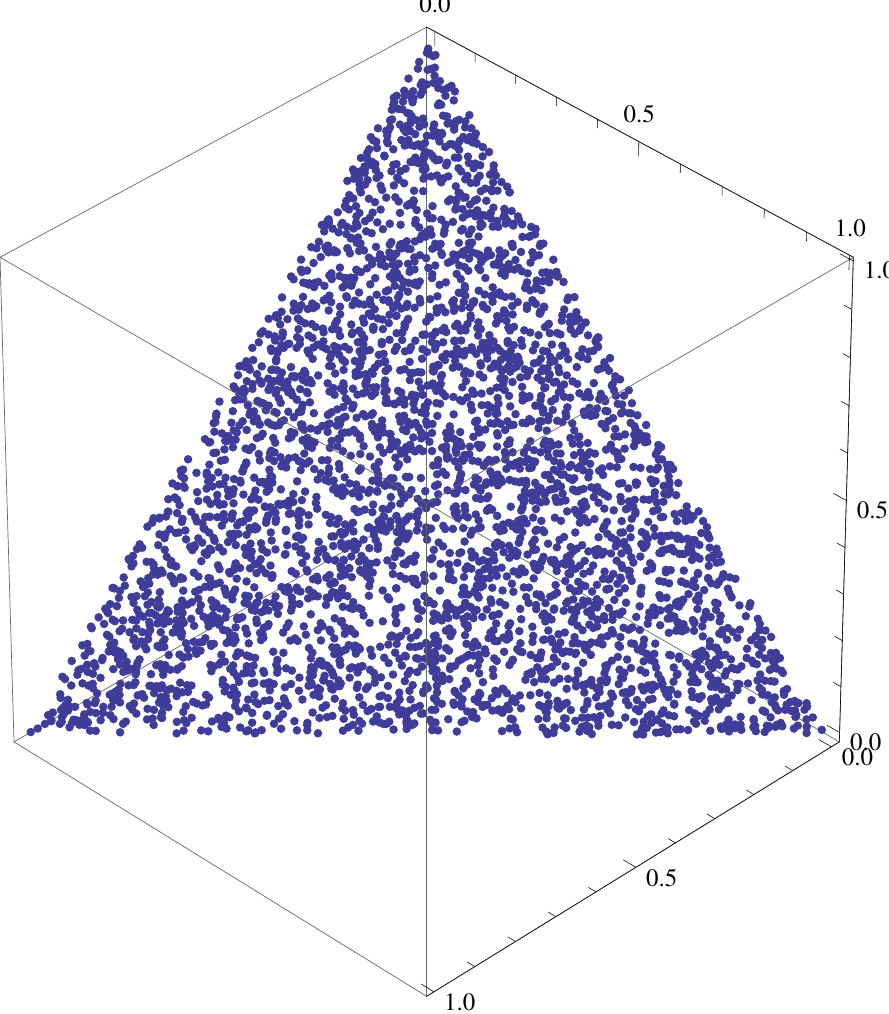}
\end{center}
\caption{Simple unbiased simplicial sample based on gaps}\label{unbiased}
\end{figure}

\subsection{Even simplicial sampling method 2}
Can we avoid the sort? We use the same concept as before but think about drawing the points directly in order. The largest sample is drawn directly from the maximum distribution of $k$ uniform samples on $(0,1)$. Then if this is the random variable $x_{Max}$, we draw the next biggest sample from the distribution of the maximum of $k-1$ uniform samples on [0,$x_{Max}$], and so on. The CDF of the $p$-maximum on the unit interval is easily seen to be $x^p$, so the quantile of the $p$-maximum on $(0,y)$ is $y*u^{(1/p)}$. So we work our way down using the quantile to generate samples from other uniform variables. The gaps are then used as before and the same picture is obtained. The author's own experience of this method is that it is slower than the first one, but this seems to be a function of the relative inefficiency of working out many fractional powers compared to a simple sort. The maturity of sort routines seems to ensure that method 1 wins over method 2\footnote{This was tried in Mathematica V7 - it is possible that other balances may be found in different computing environments}.

\subsection{Even simplicial sampling method 3}
This uses the idea that we take
\begin{equation}
w_i = \frac{Z_i}{\sum_{j=1}^N Z_j}\ ,
\end{equation}
but chooses the distribution of the $Z_i$ much more intelligently. The $Z_i$ are taken from a one-sided exponential distribution with density 
\begin{equation}
f(x) = \lambda e^{-\lambda x}
\end{equation}
for any choice of positive $\lambda$. An elementary computation with exponential and gamma distributions reveals that the CDF of e.g. $w_1$ is given by
\begin{equation}
F(x) = 1 - (1-x)^{N-1}
\end{equation}
which is the CDF of the minimum of $N-1$ points. This is just what is required when we compare with method 2. The simulation of one-sided exponentials is trivial as there is an explicit quantile function. Working it out the value of $\lambda$ drops out and we can lose some minus signs and simplify to
\begin{equation}
w_i = \frac{\log U_i}{\sum_{j=1}^N \log U_j}\ ,
\end{equation}
where the $U_i$ are uncorrelated uniform random variables on $(0,1)$. The view of the author is that this is the method of choice if you want to sample a simplex or subject of it evenly - the issue of whether such even sampling is optimal is a separate matter and we turn now to the testing of this.

\section{First test problems}
In order to develop a basic understanding of the Monte Carlo portfolio sampling and its application we first apply it to some simple QP problems. This has the advantage that we can establish the basic feasibility of the approach and highlight any difficulties. In this case we do not need to couple to samples of the return distribution, as the risk measure is a trivial function of the simulated weights. This allows us to check basic issues, but will of course leave open the question of computational feasibility of the system when both the portfolios and the distributions are sampled simultaneously.

We first consider again the 3-dimensional pure variance-minimization problem introduced by the covariance of Eqn.~(\ref{cova}). The solution to the problem\footnote{One can do this one analytically with Lagrange multipliers taking care of the boundary, and the same answers are obtained numerically by direct application of the {\tt FindMinimum} function in {\it Mathematica} 7, or indeed by the use of M.J.D. Powell's TOLMIN algorithm\cite{tolmin}. } is given by the following formula for $(w_1, w_2)$, with $w_3 = 1-w_1-w_2$.
\begin{equation}
(w_1, w_2) = \threepartdef
 {(\frac{225-120 r}{289-240 r},1-\frac{225-120 r}{289-240 r})} {r<-0.383782}   
 {(\frac{5 (48 r-125)}{576 r^2+240 r-1001},\frac{9 (40 r-17)}{576 r^2+240 r-1001})} {-0.383782 \leq r \leq0.425}  
 {(0.657895, 0)} {r > 0.425} \end{equation}
Some sample values are given in the following table.
\begin{table}[hbt]
\begin{center}
\begin{tabular}{|r|l|l|l|l|l|l|l|} \hline
 $r$ & $w_1$ & $w_2$ & $w_3$ \\ \hline
 $-0.5$ & $0.696822$ & $0.303178$   & $0.0$ \\ \hline
$0.0$ & $0.624376$ & $0.152847$   & $0.222777$ \\ \hline
$+0.5$ & $0.657895$ & $0.0$   & $0.342105$ \\ \hline
\end{tabular}
\end{center}
\caption{Sample weights for $N=3$ optimization}\end{table}
Now we write down some very simple code in {\it Mathematica} to solve the problem by the Monte Carlo method and the even simplicial sampling algorithm based on the exponential distribution. Here is the code:
\begin{verbatim}
cov3[r_] := {{64, 120 r, 25}, {120 r, 225, 50}, {25, 50, 100}};
{cova, covb, covc} = {cov3[-1/2], cov3[0], cov3[1/2]};
testa[mat_] := 
  AbsoluteTiming[rawa = Log[RandomReal[{0, 1}, {10^5, 3}]]; 
   triala = Map[#/Apply[Plus, #] &, rawa]; 
   First[SortBy[Map[{#.mat.#, #} &, triala], First]]];
Map[testa[#] &, {cova, covb, covc}]

\end{verbatim}
and this returns the output:
\begin{equation}
\left(
\begin{array}{cc}
 0.803038 & \{26.4108,\{0.693656,0.306265,0.0000786956\}\} \\
 0.803102 & \{45.5295,\{0.623704,0.152807,0.223489\}\} \\
 0.808720 & \{50.6587,\{0.656182,0.000037375,0.343781\}\}
\end{array}
\right)
\end{equation}
This suggests that the method is basically sound, taking $0.8$ seconds on a 2.8GHz machine to produce acceptable answers. We will return to prompted questions such as the precision of the zero-weight solutions, but the reader may wish to satisfy themselves that without the application of the log function the results are not as good with the same number of samples, and the computation typically still takes about $0.79$ seconds.  

\subsection{Explicit Parallelizability}
Even with such a simple test problem, we can exhibit the trivial way the solution may be parallelized. It is especially easy with {\it GridMathematica}, where one merely needs to launch slave kernels, distribute the relevant definitions, and ask each slave to find its own optima, then take the best of the best. The code is a trivial extension of that used previously, and in the following example we exhibit the result from $8$ million portfolios considered on an 8-core machine:

\begin{verbatim}

LaunchKernels[];
DistributeDefinitions[covc];
AbsoluteTiming[parallelruns =  ParallelEvaluate[
    raw = Log[RandomReal[{0, 1}, {10^6, 3}]];
    triala = Map[#/Apply[Plus, #] &, raw];
    First[SortBy[Map[{#.covc.#, #} &, triala], First]]];];
First[Sort[parallelruns]]

\end{verbatim}
which returns output, e.g. from one run of around 12 seconds, of 
\begin{equation}
\{50.6582,\{0.656187,8.4297\times 10^{-7},0.343812\}\}
\end{equation}
We will be able to do rather better still, as we shall see later - the point here is that best-of-the-best parallelization is easy. 

\subsection{General non-linear constraint management by rejection}
In practice the constraint sets are more complicated. If we stay within the long-only fully invested framework but impose further inequality conditions the simulation remains very straightforward. We simply reject those random samples not satisfying the constraints. Constraints might arise for a variety of reasons. We might want a portfolio ``beta'' to lie between certain limits, in order to constrain the portfolio compared to index behaviour. We also might wish to impose lower or upper bound conditions on individual weights, or have the total weight in a given market sector constrained to be above or below a certain level. Such inequality conditions are well suited to rejection. Equality constraints either have to be relaxed to a tight pair of inequalities to apply this method, or treated by a different method altogether. Rather than demanding $\beta=1$, for example, we might reject configurations unless they satisfy $0.95< \beta<1.05$.

With our working example, suppose we demand that in addition
\begin{equation}
w_1 \geq \frac{1}{3}\ ,\ \ \ w_2 + 1.1 w_3 \geq \frac{1}{2}\ .
\end{equation} 
To treat this we just throw out the samples that do not satisfy these conditions and optimize over those that do. Note that there is {\it no requirement that the constraints be linear}. In {\it Mathematica} the rejection may be achieved by a trivial application of the {\tt Select} command. Here we do a single-core example
\begin{verbatim}

AbsoluteTiming[rawa = Log[RandomReal[{0, 1}, {10^5, 3}]]; 
 triala = Map[#/Apply[Plus, #] &, rawa];
 constrained = 
  Select[triala, (#[[1]] >= 1/3 && #[[2]] + 1.1 #[[3]] > 1/2) &]; 
 First[SortBy[Map[{#.cova.#, #} &, constrained], First]]]
 
\end{verbatim}
which returns the outcome
\begin{equation}
\{0.853075,\{32.2394,\{0.518711,0.293995,0.187294\}\}\}
\end{equation}
We note that the calculation takes marginally longer, and compares favourably with the direct QP optimization:

\begin{verbatim}

QPweightsB[r_] := FindMinimum[{{x, y, z}.cov3[r].{x, y, z}, 
x >= 0, y >= 0, z >= 0, x + y + z == 1, 
 x > 1/3, y + 1.1 z > 1/2}, {x, y, z}];
 QPweightsB[-0.5]
\end{verbatim}
\begin{equation}
\{32.2379,\{x\to 0.518873,y\to 0.292396,z\to 0.188731\}\}
\end{equation}
The imposition of a complex set of constraints is clearly straightforward, at the price of discarding rejected samples. Note further that in this last example we have exhibited the value of the optimal objective function from both the QP solution and the Monte Carlo case, and they agree to four significant figures. This is an illustration of the notion that considering the best of a large number of simulated portfolios may be ``good enough'' to get close to optimal risk. While we will see shortly that we can do better still on the precision of the answer, it is defensible from the point of view of providing investment advice to argue that one has considered a very large number of cases and is presenting the best result from that large set. 

\subsection{More realistic problems: a 6D example}
We now increase the dimension to $N=6$ and attempt to break the method by simulating many covariance matrices and exploring the outcome.  The outcome of these studies will be condensed to exhibiting a nicely pathological example that shows we need to work harder still to make this approach viable. We will study the following positive definite covariance matrix:
\begin{equation}
\left(
\begin{array}{cccccc}
 0.0549686 & 0.144599 & -0.188442 & 0.0846818 & 0.21354 & 0.0815392 \\
 0.144599 & 1.00269 & -0.837786 & 0.188534 & 0.23907 & -0.376582 \\
 -0.188442 & -0.837786 & 1.65445 & 0.404402 & 0.34708 & -0.350142 \\
 0.0846818 & 0.188534 & 0.404402 & 0.709815 & 1.13685 & -0.177787 \\
 0.21354 & 0.23907 & 0.34708 & 1.13685 & 2.13408 & 0.166434 \\
 0.0815392 & -0.376582 & -0.350142 & -0.177787 & 0.166434 & 0.890896
\end{array}
\right)
\end{equation}
This matrix has been selected from a large number of randomly-generated ones\footnote{In fact we simulate Cholesky decompositions and then square up to recover valid covariance structures - the issue here is not the possibility of zero or negative eigenvalues that may arise through trader interference, but edge-vertex issues.} The optimum weights are given by
\begin{equation}
(0.883333,0,0.11667,0,0,0)
\end{equation}
and we note that four weights are zero and most of the portfolio is concentrated in one of the two remaining assets.  

The even simplicial sampling scheme introduced so far struggles to capture this optimum. For example, in an $8 \times 10^5$ parallel run on an 8-core machine, an example of the output is 
\begin{equation}
\{0.710654,0.0596266,0.156866,0.00340821,0.00300641,0.0664388\}\ ,
\end{equation}
and with $8 \times 10^6$ samples we only arrive at
\begin{equation}
\{0.82345,0.0141151,0.125324,0.000214034,0.00184072,0.0350566\}\ .
\end{equation}
This raises a new and interesting question over the three even-sampling schemes. They manifestly sample the simplex in an even fashion, but this is actually not always what is needed. If you look at the marginal distribution of any one of the weights, for example, in the gap approach this is the distribution of the minimum, and it is strongly biased towards zero (the bias towards zero increases as $N$ increases). This means that it is very hard to find configurations where one or two weights have a high value. A related and more subtle problem is that although we are probing the interior in an even manner, the optimum above has four weights that are essentially zero, so that we need to think harder about configurations on a boundary, and to do so in a way that will not compromise finding interior optima that are, e.g., close to $1/N$. 
\section{Changes of measure: Edge-Vertex Biased schemes}
The methods for evenly-sampling the interior do not readily capture the edge and vertex solutions, as has been made clear by the $N=6$ test. We need some method of changing the measure so as to also capture solutions along the edges and vertices. In approaching this one can consider a theoretical approach, e.g. trying to optimize the capture of both interior and boundary points, among a class of methods such as
\begin{equation}
w_i = \frac{f(U_i)}{\sum_{j=1}^N f(U_j)} \ .
\end{equation}
One might also consider separate sampling of faces, edges and vertices at the price of some combinatorial complexity.

\begin{figure}[hbt]
\begin{center}
\includegraphics[scale=0.7]{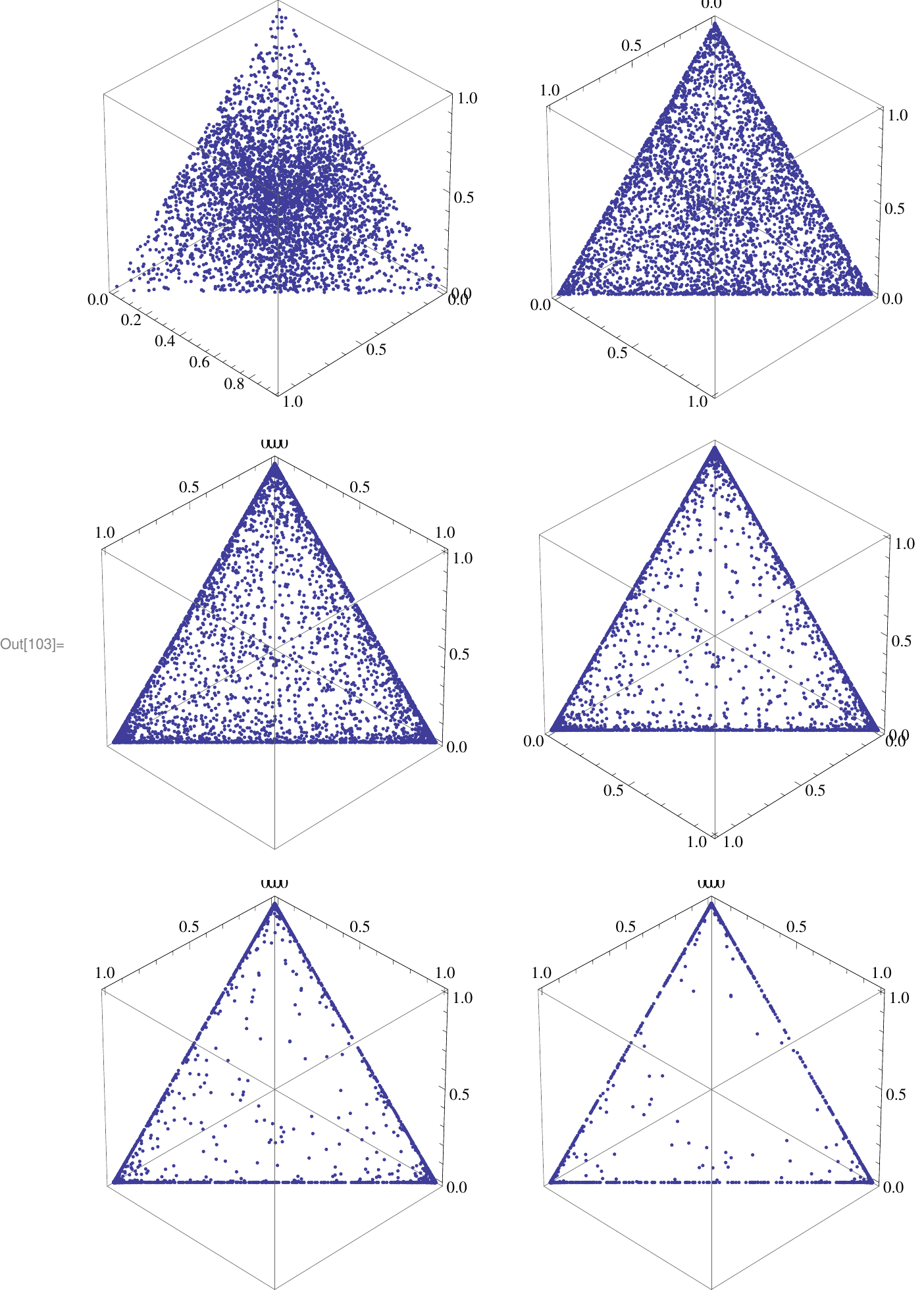}
\end{center}
\caption{Sequential edge-vertex biasing of samples}\label{sequential}
\end{figure}

However, it seems that a very effective solution can be obtained by combining a little insight from the theory of norms with some numerical experimentation. If we consider the $L_q$ norm on a vector $V_i$ in the form
\begin{equation}
||V||_q = \left(\sum_{i=1}^N V_i^q\right)^{1/q}\ \ ,
\end{equation}
we know that as $q\rightarrow \infty$ then
\begin{equation}
||V||_q \rightarrow \max_i V_i\ \ .
\end{equation}
In the current context we can adapt this idea by observing that, except in the case of two or more $U_i$ being identical, the quantity
\begin{equation}
\frac{U_i^q}{\sum_{j=1}^N U_j^q} .
\end{equation}
will approach, as $q \rightarrow \infty$ a vector of all zeroes except for a one in the slot corresponding to the largest $U_i$.  Intermediate values of $q$ will have intermediate concentrations of entries in the higher entries. In order to keep matters computationally trivial, we will consider $q = 2^p$ for $p=0,1,2\dots$, so that the bias evolves by elementary squaring. In Figure (\ref{sequential}) we consider $p=0,1,2,3,4,5$ so that we go to to $q=32$ based on 2000 initial points.

\section{Second test phase}
We now return to the testing of the modified scheme. We can re-investigate both our $N=3$ and $N=6$ example without or with any parallelization. We will focus attention on the pathological $N=6$ case, as it is the location of boundary optima which concerns us. The idea is to 

\begin{itemize}
\item create $k$ samples $U_{lm}$, $l=1,\dots N$ $m=1,\dots,k$ from the $N$-dimensional hypercube,
\item form all their iterated squares $U_{lm}^{2^p}$ up to some power (e.g. $32 = 2^5$),
\item form all the simplex samples
$$
w_{l,mp} = \frac{U_{lm}^{2^p}}{\sum_{j=1}^N U_{jm}^{2^p}} \ \ ,
$$
\item Minimize the risk function over all $m$ and $p$.
\end{itemize}
A sample calculation was done where we took $k=20,000$ on each of 8 processors, with iterated squaring to $32$, so that $8\times 20,000 \times 6 = 960,000$ portfolios were considered in a 2sec computation, with the weights outcome 
\begin{equation}
\{0.88333,\text{2.0E-14},0.116658,\text{3.9E-8},\text{7E-43},0.00001166\}
\end{equation}
So the problem has been solved. Fig.~\ref{sequential} makes it clear that both interior and boundary points may be found with equal ease by considering the union of all the samples in the six scatterplots.

\newpage
\section{Full Distributional Risk Measures}
Having established the feasibility of an optimization based on edge-vertex-biased sampling, for directly computable functions of the weights we now generalize to the case where the multivariate return distribution is itself simulated. What we need to do is essentially the same as before, except that rather than working out simple functions of the weights such as $w.C.w$, we have to consider functions of the form
\begin{equation}
{\rm Risk}(w, {\rm samples}, {\rm params})
\end{equation}
where the Risk function has inputs that are weights,  the entire distribution realized in terms of samples, and auxiliary parameters such as
\begin{itemize}
\item Lagrange multipliers for QP problems;
\item expected returns;
\item quantile levels for VaR, CVaR applications;
\item $p,q,b$ values for Omega, Sortino;
\item parameters of utility or prospect functions;
\item $\dots$.
\end{itemize}
In principle we just compute the samples in advance and send them into the computation as a big extra matrix parameter, distributing them to all the cores in the case of a parallel computation.

\subsection{Consistency, stability and sample size}
In moving to a characterization of risk through a {\it sampled} distribution we must tackle an important issue that arises whenever a multivariate system is realized through sampling\footnote{Similar issues arise with basket option pricing.}.  The volatility and dependency structure may be characterized in various ways, but for the sake of explaining this idea we will assume that a covariance has been defined from some data.\footnote{It could equally well be the parameter of a chosen Archimedean copula together with some variances.} We call this the {\it input covariance.} We now sample the distribution using these parameters and then recompute the {\it realized covariance}. The realized covariance will not be the same as the input covariance, though we expect that the two will get closer in some sense as the number of samples increases. 

Given that, e.g. in the case of QP, the computed optima may jump around as a function of the dependency data supplied, this needs careful management. I do not have a solution to this beyond suggesting that the number of return samples employed should be as large as is practical, and that the stability of the optimal weights as the number of return samples is increased should be considered. In general up to five distinct computations might be possible, and all five are possible in the case of a QP problem, which gives scope for working out ``how much is enough'':

\begin{enumerate}
\item Analytic QP on the input covariance;
\item Analytic QP on the realized covariance;
\item Monte Carlo Edge-vertex biased sampling of the simple analytic function (as above) with input covariance;
\item Monte Carlo Edge-vertex biased sampling of the simple analytic function (as above) with realized covariance;
\item Monte Carlo Edge-vertex biased sampling with full distributional return sampling, implicitly with realized covariance.
\end{enumerate}

This might seem somewhat burdensome, but we are just looking for reasonable agreement between computations 1 and 3, and/or 2 and 4 in order to be sure that we have a sufficient number of random portfolios, and then agreement between 1 and 2, 4 and 5 to know we have enough samples of the return distribution. While some analytic estimates of when this works are desirable, I will continue in the rather pragmatic experimental spirit of this paper in trying to find out what works. I will continue with the pathological $N=6$ example where we have done the work of comparing 1 and 3, so from here on it is all about the number of samples in the return distribution. Some example results for calculation 1 vs 2 are given in the following table, using {\it Mathematica's} {\tt FindMinimum} function. The norm function $\Delta$ is given by
\begin{equation}
\Delta = {\rm Max}_{i,j} |C_{ij}(realized) - C_{ij}(input)|
\end{equation}
and the zero weights are omitted, all being small in the examples computed.
\begin{table}[hbt]
\begin{center}
\begin{tabular}{|r|l|l|l|l|l|l|l|} \hline
 $k$ & $\Delta$ & $w_1$ & $w_3$ \\ \hline
  $100$ & $0.2721$ & $0.877$   & $0.123$ \\ \hline
 $1000$ & $0.077$ & $0.880357$   & $0.116333$ \\ \hline
$10000$ & $0.01976$ & $0.883597$   & $0.116403$ \\ \hline
$100000$ & $0.10102$ & $0.883168$   & $0.116832$ \\ \hline
${\rm input}$ & $0$ & $0.883333$   & $0.11667$ \\ \hline
\end{tabular}
\end{center}
\caption{Covariance and weights stability as function of return sample size, exact QP}\end{table}

Of course, this is all Monte Carlo, so recomputation causes this table to change, but our general experience is that $10,000$ samples is a minimum safe number. Going beyond this helps the solution precision, and is of course necessary if one wishes to probe VaR or CVaR measures deep into the tail. 

We now jump straight to calculation type 5, where both the portfolios and the return distribution are simulated. This is our first example of ``The Full Monte''.  The computation of the return distribution samples is done in advance, and results in a $10000 \times 6$ matrix given the symbol {\tt mixed}.  The calculation is shown as a {\it Mathematica} session using some prototype packages and built-in simulation tools to do the work. First we define a multivariate uncorrelated normal sampler, a random Cholesky matrix simulator, seed the random number generator to reproduce the pathological case under consideration and then create the correlated samples.
\begin{verbatim}

rawnormal[n_, dim_] := RandomReal[NormalDistribution[0, 1], {n, dim}]

SeedRandom[20]

ranchol[n_] := 
 Table[If[i > j, 0, RandomReal[{-1, 1}]], {i, 1, n}, {j, 1, n}]
 
six = ranchol[6];
 
rawb = RandomReal[NormalDistribution[0, 1], {10000, 6}];
 
mixed = Map[#.six & , rawb];

\end{verbatim}
At this stage we have $10000$ $6$-vectors with the realized covariance. Our package consists of a suite of prototype risk functions. A load and query command pair
\begin{verbatim}

Needs["PMCOptimizer`RiskFunctions`"]
?PMCOptimizer`RiskFunctions`*

\end{verbatim}
outputs a list of functions containing objects like {\tt ConditionalValueAtRisk, Omega, ValueAtRisk, MeanVariance, NegativeSharpeRatio} and so on. A second component of the package contains the optimizers developed so far based on this approach, which at this stage has a developed form of the long-only problem optimizer with no auxiliary constraints (the addition of further constraints is not difficult and is being functionalized). 
\begin{verbatim}

Needs["PMCOptimizer`Optimizers`"]

?PMCOptimizer`Optimizers`*

\end{verbatim}
returns the information:
{\it LongOnlyPMCOptimizer[samples, numberofassets, 
basesearchsize, evbiasdepth, riskfunction, riskfunctionparameters] 
minimizes the supplied risk function with given 
riskfunctionparameters, based on an input of terminal multi-asset 
samples for numberofassets assets, using a basesearchsize 
and a defined iterative vale of evbiasdepth.

E.g. of use 
LongOnlyPMCOptimizer[mixed, 6, 1000 5, MeanVariance, 0.0] 

Does classic mean-variance optimization with zero return tilt for a
sample defined by the N by 6 matrix mixed, with 1000 basic simplex 
samples iteratively biased five times. The method is parallelized.}

So now we just have to run it using the command (the seeding is to produce reproducible random results from a bunch of slave kernels - in live applications one would probably do this to avoid ``turnover'' generated by Monte Carlo noise.)
\begin{verbatim}

ParallelEvaluate[SeedRandom[$KernelID*100]];
LongOnlyPMCOptimizer[mixed, 6, 1000, 6, MeanVariance, 0.0] 

\end{verbatim}
This returns the result
\begin{equation}
\{9.983915,\{0.0274786, 
\end{equation}
\vspace{-0.35in}
$$
\{0.882869, 1.22605*10^{-14}, 0.117122, 9.9529*10^{-7}, 1.2674*10^{-15}, 8.05487*10^{-6}\}\}\}
$$
which is a 10 second computation on my deskside 2008 Mac Pro 8-core. 

The author is happy to stipulate that this is possibly the least efficient QP algorithm yet written down. The upside is that now we can replace the function {\tt MeanVariance} with {\it any function of the distribution of portfolio returns} and the speed of the method will scale easily inversely with the number of cores deployed, provided that each core can store the return distribution and do the basic computations and sorting. Furthermore the precomputation of the distribution needs only to be done once and can be based on any multivariate structure, empirical data, or the result of evolving some complicated stochastic model. This latter feature is of course in common with the RU-method \cite{cvaropt} but the ability to swap in any risk function is a novel one, and the size of the return distribution can be increased provided each core has sufficient memory and the sort time does not become prohibitive. 
 
\subsection{Example: VaR/CVaR/Sharpe Ratio/Omega optimization}

To illustrate the flexibility of this method, we now replace the {\tt MeanVariance} function with other functions of the full distribution, again with controlled seeding. This is a somewhat vacuous test however, as for multivariate zero-mean Gaussian we would expect the same optimal weights to be obtained for corresponding calculations, provided enough samples are taken. For example, both the VaR and CVaR are obviously minimized by making the portfolio distribution as narrow as possible, which would reduce to pure variance minimization. So the similarity in the following results for multivariate Gaussian is in fact a useful check that we are doing the right computations. 

\begin{verbatim}

In[45]:= ParallelEvaluate[SeedRandom[$KernelID*100]];
In[46]:= LongOnlyPMCOptimizer[mixed, 6, 10^3, 6, ValueAtRisk, 0.025] 

Out[46]= {12.210430, {0.327577, {0.871788, 4.58122*10^-30, 0.128205, 
   1.40685*10^-6, 5.2755*10^-6, 3.54997*10^-8}}}

In[49]:= ParallelEvaluate[SeedRandom[$KernelID*100]];
In[50]:= LongOnlyPMCOptimizer[mixed, 6, 
 10^3, 6, ConditionalValueAtRisk, 0.025] 

Out[50]= {12.627039, {0.393196, {0.882869, 1.22605*10^-14, 0.117122, 
   9.9529*10^-7, 1.2674*10^-15, 8.05487*10^-6}}}

In[51]:= ParallelEvaluate[SeedRandom[$KernelID*100]];
In[52]:= LongOnlyPMCOptimizer[mixed, 6, 10^3, 6, MeanVariance, 0.0] 

Out[52]= {9.920736, {0.0274786, {0.882869, 1.22605*10^-14, 0.117122, 
   9.9529*10^-7, 1.2674*10^-15, 8.05487*10^-6}}}

In[53]:= ParallelEvaluate[SeedRandom[$KernelID*100]];
In[54]:= LongOnlyPMCOptimizer[mixed, 6, 
 10^3, 6, NegativeSharpeRatio, -1.0] 

Out[54]= {9.944519, {-6.01598, {0.882869, 1.22605*10^-14, 0.117122, 
   9.9529*10^-7, 1.2674*10^-15, 8.05487*10^-6}}}

In[55]:= ParallelEvaluate[SeedRandom[$KernelID*100]];
In[56]:= LongOnlyPMCOptimizer[mixed, 6, 10^3, 6, NegativeOmega, -0.5] 

Out[56]= {64.089293, {-4169.27, {0.889907, 1.08026*10^-6, 0.109982, 
   6.87834*10^-9, 3.13299*10^-17, 0.000109915}}}

\end{verbatim}
Changing the nature of the input distribution {\it mixed} clearly will lead to a richer set of results, but the results above may also be regraded as a testament, if any were needed, to the substance of Markowitz's original insight. 

\subsection{Example: CVaR optimization vs Rockafellar-Uryasev}
Considerable interest has been stimulated in CVaR optimization in part due to its coherence properties but mainly due to the development of an ingenious solution technique in \cite{cvaropt}. We can use this as a further comparison check. Rockafeller and Uryasev considered, as a test problem, the optimization of a portfolio consisting of three assets with covariance given by
\begin{equation}C_{ij} = 
\left(
\begin{array}{ccc}
 0.00324625 & 0.00022983 & 0.00420395 \\
 0.00022983 & 0.00049937 & 0.00019247 \\
 0.00420395 & 0.00019247 & 0.00764097
\end{array}
\right)
\end{equation}
and expected returns given by
\begin{equation}
R_i = \{0.010111,0.0043532,0.0137058\}
\end{equation}
They analyzed the problem in various forms, but here we will focus on the CVaR optimization results presented in Table 5 of \cite{cvaropt}. This is an interesting comparison as the returns, as here, are represented by direct simulation, so we can do a reasonable like-with-like comparison between their LP approach and the proposed random portfolio method.  We must of course bear in mind the 10 year difference in computer speed. The were working on a 300 MHz machine whereas the results here are at 2.8GHz, with optional multi-core technology. 

This part of the work of \cite{cvaropt}, in common with our own study, makes a comparison with traditional QP based on Gaussian statistics, based on the fact that the optimal CVaR, if done correctly, should coincide with the simple minimum variance portfolio. Rockafellar and Uryasev find the optimal weights under the minimum variance approach to be
\begin{equation}
\{0.452013, 0.115573, 0.432414\}
\end{equation}
with the resulting minimum variance of 0.00378529 and the CVaR results being
\begin{equation}
\{0.096975, 0.115908, 0.152977\}
\end{equation}
and the three considered quantiles of $0.1, 0.05, 0.01$ respectively.  A quick application of {\tt FindMinimum} in {\it Mathematica} produced an identical optimal variance and optimal weights agreeing to 5 sig. figs. 

For the direct Monte Carlo CVar optimization the parameters employed are: $J=1000, 3000, 5000, 10,000, 20,000, 100,000$ for the return distribution size. Our previous experiments suggests that the results to take seriously are those with $J \geq 10,000$ and a good number of sample portfolios. We consider base portfolio sample sizes based on our early verification, so we have $20000$ base portfolios iterated to the 32nd power, on each of 8 kernels, so that 960,000 random portfolios are considered. In the light of the results being close to equal weight, we could consider doing a recomputation with less EV-bias and more base portfolios, but would prefer to leave this out here - it might appear to be using some foresight, and in any case the matter of intelligent refinement is a subject of current research. The quantiles considered are, as with the RU paper, $0.1, 0.05, 0.01$. 

\begin{table}[hbt]
\begin{center}
\begin{tabular}{|l|r|l|l|l|l|l|r|} \hline
Q'tile&$J$& $w_1$ & $w_2$ & $w_3$ & CVaR & dCVaR & Time  \\ 
$1-\beta$ &&  &   &  & optimal  & opt-QP & {\it secs}  \\ \hline
$0.1$ & $1000$ & $0.444912$  &$0.118273$   & $0.436815$  &$0.0948004$  &$-0.0021746$& $5.1$  \\ 
$0.1$ & $3000$ &$0.516779$&$0.0906586$&$0.392562$&$0.0971534$&$+0.0001784$&$17.7$ \\ 
$0.1$ & $5000$ &$0.415496$&$0.129594$&$0.45491$&$0.0992909$&$-0.0870459$& $34.7$  \\ 
$0.1$ & $10000$ &$0.457608$&$0.113378$&$0.429014$&$0.096326$&$-0.000659$&$68.4$ \\ 
$0.1$ & $20000$ &$0.441418$&$0.119613$&$0.438969$&$0.0976096$&$+0.0006346$&$140.9$  \\ 
$0.1$ & $100000$ &$0.45274$&$0.115244$&$0.432016$&$0.0972891$&$+0.0003141$&$816.2$  \\ \hline \hline
$0.1$ &{\rm QP} &$0.452013$&$0.115573$&$0.432414$&$0.096975$&$0$&NA  \\ \hline \hline
$0.05$ & $1000$ &$0.611977$&$0.0540555$&$0.333967$&$0.117257$&$+0.001349$&$5.5$  \\ 
$0.05$ & $3000$ &$0.383387$&$0.141948$&$0.474665$&$0.11562$&$-0.00028$&$18.5$  \\ 
$0.05$ & $5000$ &$0.451571$&$0.115731$&$0.432698$&$0.11573$&$-0.000178$&$35.0$  \\ 
$0.05$ & $10000$ &$0.462245$&$0.111638$&$0.426117$&$0.114866$&$-0.001042$&$71.6$  \\ 
$0.05$ & $20000$ &$0.477872$&$0.105622$&$0.416506$&$0.116242$&$+0.000334$&$142.5$  \\ 
$0.05$ & $100000$ &$0.462419$&$0.111555$&$0.426025$&$0.116252$&$+0.000344$&$821.7$  \\ \hline \hline
$0.05$ &{\rm QP} &$0.452013$&$0.115573$&$0.432414$&$0.115908$&$0$&NA  \\ \hline \hline
$0.01$ & $1000$ &$0.575114$&$0.0681314$&$0.356755$&$0.146881$&$-0.006096$&$5.5$  \\ 
$0.01$ & $3000$ &$0.5432$&$0.0804313$&$0.376368$&$0.150387$&$-0.00259$&$18.2$  \\ 
$0.01$ & $5000$ &$0.306275$&$0.171566$&$0.522159$&$0.15732$&$+0.004343$&$35.3$  \\ 
$0.01$ & $10000$ &$0.492627$&$0.0999324$&$0.40744$&$0.153361$&$+0.000384$&$69.9$  \\ 
$0.01$ & $20000$ &$0.503307$&$0.0958251$&$0.400868$&$0.152121$&$-0.000856$&$142.2$  \\ 
$0.01$ & $100000$ &$0.409185$&$0.131996$&$0.458818$&$0.152366$&$-0.000611$&$187.9$  \\   
$0.01$ & $200000$ &$0.46145$&$0.11193$&$0.42662$&$0.152727$&$-0.00022$&$1809.8$  \\    \hline \hline
$0.01$ &{\rm QP} &$0.452013$&$0.115573$&$0.432414$&$0.152977$&$0$&NA  \\ \hline \hline
\end{tabular}
\end{center}
\caption{Recomputation of Rockafellar-Uryasev CVaR example, 960,000 Random Portfolios}\end{table}

Inspection of the tables reveals some interesting observations. It is clear that there is significant instability in the results for $J<10,000$, confirming our earlier circumspection on having a good-sized portfolio return sample. For deep tail CVaR it is also clear, as might have been expected, that for the 0.01 quantile rather more samples are needed. The timings become awkward as we let $J$ push through $100,000$, but it must also be borne in mind that not only is this a prototype code using largely interpreted {\it Mathematica}, we are only just in the early years of multi-core computation and the times can be slashed by parallel deployment. This will be necessary, as is made clear by the one case of $J=200,000$ added for the 0.01 quantile case.  

As in the RU-study, these results can be recomputed with a variety of different simulation methodologies. RU recomputed with quasi-random sobol sequences generating the return samples. We could of course apply this notion to both our return distribution, and, given that the portfolio samples also originate on a hypercube, these might be similarly generated, though the interaction with the parallelization would need thought - the pure Monte Carlo approach parallelizes rather trivially provided each core or kernel can store the return distribution. 

Of course, we can recompute examples like this with {\it any} suitable risk function. which is the payback for the increased computation associated with the random portfolio method. 

\newpage
\subsection{Omega optimization}
The Omega function was introduced by Cascon, Keating and Shadwick (CKS)\cite{omega}. While introduced independently, with hindsight it may be viewed as a special case of the variability ratios introduced earlier. If we consider
\begin{equation}
\Phi(b,1,1) \equiv \Omega(b) =  \frac{E[(X-b)^+]}{E[(b-X)^+]}
\end{equation}
we obtain an object which is superficially similar to the price of a Call option struck at $b$ divided by a corresponding Put. This is a slightly misleading viewpoint as the expectations have to be taken in the real-world measure. Writing out the expectations explicitly in terms of a pdf $f(x)$ and integrating by parts leads to the form originally written down by CKS:
\begin{equation}
\Omega(b) = \frac{\int_b^{\infty}(1-F(x))}{\int_{-\infty}^b F(x) dx}
\end{equation}
where $F(x)$ is the CDF (cumulative distribution function) with $F'(x) = f(x)$. For the financial motivation see \cite{omega}. 

These is an interesting objective function, and its optimization is consistent with some traditional approaches in a Gaussian environment. Let X have a distribution that is Gaussian with mean $\mu$ and standard deviation $\sigma$. Let the standard normal density function be $\phi(x)$ and CDF $\Phi(x)$, with $\Phi'(x) = \phi(x)$. Then some algebra and calculus gives us
\begin{equation}
\Omega(b) = \frac{\phi(z) + z \Phi(z)-z}{\phi(z) + z \Phi(z)}
\label{omsimp}
\end{equation}
where 
\begin{equation}
z = \frac{b-\mu}{\sigma}
\end{equation}
We can check that the RHS of Eqn.~(\ref{omsimp}) is a strictly decreasing function of z. Hence the maximization of $\Omega$ in a Gaussian model is simply the minimization of $(b-\mu)/\sigma$ for the portfolio. This is just the minimization over allowed choices of weight, of the negative of a Sharpe ratio with benchmark the threshold $b$ instead of the risk-free rate. In general, with other distributions with possible non-trivial skewness and kurtosis, we get something else, and its optimization in general is less straightforward.

The optimization of $\Omega$ over possible portfolio weights has been considered by a few writers  in the last decade.  Kane {\it et al} \cite{nagomega} considered an optimization based on empirical data for three assets. One approach they took was a Nelder-Mead algorithm, well adapted to problems such as this where the gradient is not readily specified, and other methods were used for higher-dimensional problems. Unfortunately, not all the properties of the input data are explicitly specified in this work so we are not able to do a detailed cross-check to see if we recover the same outputs. Mausser {\it et al} \cite{risklpomega} describe an elegant transformation of the optimization to a problem in linear programming, but only summarize the moment statistics of the marginals, making a detailed benchmarking exercise against their results more complicated.   Passow \cite{passow} gives an elegant analytical treatment in terms of Johnson distributions for the portfolio, but again is not fully explicit about the nature of the dependency structure.\footnote{Johnson distributions are a flexible method of modelling the non-Gaussian aspects of financial returns, but in the view of this author at least, the common approach of calibrating them using skewness and kurtosis is problematic in the light of recent work exposing the heavily fat tailed nature of asset returns. See e.g. \cite{platent,stanley} for evidence of quartic and cubic power law decay for asset returns, rendering kurtosis (and possibly skewness in the light of \cite{stanley}) calibration highly unstable. This is a separate matter to the issue of $Omega$ optimization, but needs to be noted.} A more explicit example has been constructed by Avouyi-Dovi, Morin and Neto (ADMN for short) \cite{clearomega}, who characterize their model completely in terms of a T-copula dependency model and GARCH evolution on the marginals. 

In view of the complications on explicitness or model complexity in published work\footnote{The author would be grateful to hear of a published clear example with all aspects of marginals and dependency made explicit in a simple model.}, we will propose a simpler benchmark for comparing $\Omega$ optimizers, based on a simplification of the model in \cite{clearomega}. This will hopefully be much more accessible to others for cross-testing with no ambiguities about parameters and no modelling threshold other than having a candidate $\Omega$-optimizer. We aim to revisit the full analysis of \cite{clearomega} in later work.

\subsection{Simplified ADMN Omega optimization model}
In the ADMN model a portfolio of three indices $(A,B,C) = $ (Germany, UK, US) is considered. In our simplified model the problem is described by the correlation matrix (the unconditional $\rho$ from Table C.1 of \cite{clearomega})

\begin{equation}\rho_{ij} = 
\left(
\begin{array}{ccc}
 1.00000000 & 0.47105463 & 0.35635569 \\
 0.47105463 & 1.00000000 & 0.44091699 \\
 0.35635569 & 0.44091699 & 1.00000000
\end{array}
\right)
\end{equation}
In the ADMN paper a clear statement of the degrees of freedom of the T-copula is not, so far as we can tell, given, but the noise structure of the marginals appears to have a d.o.f. around $\nu=9$, so our simplified model will take a dependency structure given by a standard  multivariate T itself with a single value for the degrees of freedom (in full generality we might expect to have a copula d.o.f., and if the marginals were also T, three marginal d.o.fs.). That leaves us specifying asset vols and means. For the means we will take the $\mu$ parameter from Table C.2 of \cite{clearomega}:
\begin{equation}
\mu_i = \{0.18963989, 0.16829560, 0.2788619 \}
\end{equation}
and for the standard deviations we will take the values also from C2. 
\begin{equation}
\sigma_i = \{2.3251341, 2.0430214, 1.8134084 \}
\end{equation}
These values are given in \cite{clearomega} as being those pertinent to the holding period, so hopefully constitute relevant parameters for when the GARCH details
are washed out in this simplified form. 

The VaR parameters of the marginals were given in \cite{clearomega} and we can compare the values cited there in Table C2 with those from our simplified $T_9$ multivariate model. These can be worked out from the formula, for general d.o.f. $n$, for the signed VaR (negative numbers denote a loss)
\begin{equation}
{\rm VaR}_i(u) = \mu_i + \sigma_i \sqrt{\frac{(n-2)}{n}} F_{\nu}^{-1}(u)
\end{equation}
where $F_{\nu}^{-1}(u)$ is the inverse CDF or quantile function for a Student T distribution with $\nu$ d.o.f. This was written down by Shaw \cite{shawt} as
\begin{equation}F_n^{-1}(u) = \text{sgn}\left(u-\frac{1}{2}\right)
\sqrt{n \left(\frac{1}{I_{\text{If}\left[u<\frac{1}{2},2 u,2 (1-u)\right]}^{-1}\left(\frac{n}{2},\frac{1}{2}\right)}-1\right)} 
\end{equation}
and the further correction of $\sqrt{\frac{(n-2)}{n}}$ is due to a {\it standard} T distribution having variance $n/(n-2)$. 

\begin{table}[hbt]
\begin{center}
\begin{tabular}{|l|l|l|l|} \hline
 & Asset A& Asset B& Asset C \\ 
$u$ & Var($u$)& Var(u)& Var(u) \\ \hline 
$0.05$ & -3.5693 & -3.13456 & -2.65279 \\
$0.01$ & -5.59593 & -4.9153 & -4.2334 \\ \hline
\end{tabular}
\end{center}
\caption{VaR values for simplified Omega model, $T_9$ distribution}\end{table}

The MC-POPE model is run for $J=50,000$ return realizations for various values of the threshold. The computation of Omega is slightly more demanding so that number of random portfolios sampled was taken at a lower figure of  $2000 \times 6 \times 8 = 96000$, realized as $2000$ samples on each of 8 processors with 6 different biasing values. The results are as shown in Table \ref{omegas}.

\begin{table}[hbt]
\begin{center}
\begin{tabular}{|r|r|l|l|l|} \hline
Threshold & Omega& Asset A& Asset B & Asset C \\ 
$b$ & optimal & $w_1$& $w_2$ & $w_3$ \\ \hline 
$-4$ & $662.7$ & $0.22$ & $0.26$ & $0.52$ \\ \hline
$-3$ & $180.0$ & $0.20$ & $0.25$ & $0.55$ \\ \hline
$-2$ & $37.4$ & $0.19$ & $0.26$ & $0.55$ \\ \hline
$-1$ & $7.9$ & $0.19$ & $0.23$ & $0.58$ \\ \hline
$0$ & $1.5$ & $0.07$ & $0.04$ & $0.89$ \\ \hline
$+1$ & $0.4$ & $1$ & $0.0$ & $0.0$ \\ \hline
$+2$ & $0.13$ & $1$ & $0.0$ & $0.0$ \\ \hline
\end{tabular}
\label{omegas}
\end{center}
\caption{Omega-optimal weights from MC-POPE and simplified ADMN model}\end{table}
The general levels of the optimal Omega and corresponding weights found from MC-POPE exhibit similar trends to those in \cite{clearomega}, despite the significant simplification of this model compared to that one, as can be verified by comparing Table \ref{omegas} here to the relevant components of Tables G1 and G2 in \cite{clearomega}. Note that \cite{clearomega} does not report results for $s >0$. There is nothing about the Monte Carlo approach requiring constraints on $b$, provided the sample extends widely enough to capture more extreme values. As reported in \cite{risklpomega}, the transformation to an LP problem in fact only works when $\Omega > 1$, so there is something essentially different about the two last cases, which we now consider.
\subsection{The range of sensible thresholds}
Let us return to the definition as a ratio of the payoffs of (real-world-measure) call and put:
\begin{equation}
\Omega(b) =  \frac{E[(X-b)^+]}{E[(b-X)^+]}
\end{equation}
We can rewrite this as
\begin{equation}
\Omega(b) =  \frac{E[(X-b)^+] - E[(b-X)^+]+E[(b-X)^+]}{E[(b-X)^+]}
\end{equation}
Now using the same kind of argument as is developed for Put-Call parity in option-pricing, the first two terms in the numerator may be rewritten, leaving us with
\begin{equation}
\Omega(b) =  \frac{E[(X-b)] + E[(b-X)^+]}{E[(b-X)^+]} = \frac{\mu_X-b}{E[(b-X)^+]}+1 \label{omnew}
\end{equation}
This form of the $\Omega$ function has been known for some time, see e.g. the presentation by Keating \cite{ckslides}. It is now clear that
\begin{itemize}
\item $\Omega > 1$ if and only if $\mu_X > b$;
\item $\Omega <  1$ if and only if $\mu_X <b$;
\item Optimization of $\Omega$ is only sensible if $b$ is no greater than the maximum return of any asset in the portfolio.
\end{itemize}
No feasible portfolio can have a return greater than the maximum return of any single asset, so inserting a threshold that is too high (while it is computable and indeed optimizable) is, in our view, financial nonsense. In practice it appears that the MC-POPE optimizer then still works, but picks the configuration with the {\it largest} downside volatility in order to {\it maximize} the denominator in Eqn.~(\ref{omnew}). In our example there is an unstable transition from an optimal weight set of $\{0,0,1\}$ when $b$ is juts below the threshold to $\{1,0,0\}$ when it is just above. The precise switch point can vary due to the fact that the $\mu_X$ is the mean of the realized distribution. In the case of Gaussian or Student-T models this can easily be fixed by using antithetic variables to se the realized mean to precisely the input mean, but this route may not be so clear for other distribution models.  In reality it is better practice to keep $b$ clearly below the maximum return on any component asset and run the optimization for a few values in this region to ensure one has not switched to a bad state. In practice one would not wish to go near the configuration with nearly everything in one asset, unless one is a ``Twain-extremist''. In a QP context trying to constrain the return to be greater than the maximum return on an asset would result in infeasability. In the case of Omega, while this function may be worked out for any threshold, trying to do an optimization for $b > Max[\mu_i]$ is possible but financially irrational.

\section{Conclusion}
We have developed a method of simplicial sampling, known as sequential edge-vertex biasing, which makes it possible to combine Monte Carlo simulation of both portfolios and returns to produce a highly flexible approximate optimizer capable of optimizing a diverse collection of risk functions. The input distribution may be varied at will and the computation parallelizes easily. We call this ``MC-POPE''.

The method makes good use of parallel computing technology. The results here have been computed on a deskside $2\times4$-core machine from 2008, so that in late 2010 a factor of two is already available on single machines, and 2-3 {\it orders of magnitude} speed improvement are possible with a cluster now, quite apart from any internal improvements that may be possible and the growth in cores per chip. 

Future possible developments include

\begin{itemize}
\item Deterministic hypercube sampling;
\item Refinement;
\item Short-selling management;
\item Robustification based on sampling uncertain parameters.
\end{itemize}

\noindent
The method is slower than traditional QP techniques when applied to traditional problems, but the method's generality and scalability with multi-core computing technology makes it an attractive option for future asset-allocation work.  

A further  intriguing possibility is that of massively accelerating these computations by the use of GPGPU boards running CUDA or OpenCL. The best way of exploiting the architecture of e.g. Quadro and GTX graphics cards to this end is under investigation.

While most of the examples presented here have been multivariate Gaussian, this is solely for the purposes of testing the new methodology against known test problems using the QP philosophy coupled to the distributions being characterized {\it only} by their mean and variance. In reality other marginal distributions and dependency structures may be employed, preferably based on sound data analysis. For example, Fergusson and Platen \cite{platent} argued that the MLE estimator of daily world index log-returns was a Student t distribution with four degrees of freedom - such marginals are easily simulated e.g. using the methods of \cite{shawt}. Various studies have captured evidence for Student, Variance Gamma, NIG distributions, and these can be simulated either in their multivariate forms, where available, or where the dependency is captured by a suitable copula. It will be clear that this variation represents a precomputation whose results are fed to the MC-POPE optimizer in just the same way as Gaussian simulations. 

The author acknowledges that this method represents almost pure brute force, but with an effectiveness sharpened by the method of edge-vertex biasing to treat this particular class of investment problems. The fact that it makes particularly good use of the current direction of evolution of computer technology is a bonus that means that it is capable of industrial-strength deployment, with a clear route to improvements in precision and/or speed.

\section*{Acknowledgments}
The author thanks Reza Ghassemieh and Roger Wilson for a helpful combination of support and scepticism. Josephine Gerken is thanked for her work on the RU method and CVaR test problem. The author is indebted to Apple for the provision of grid computing resources under the Apple Research and Technology Support Programme, to Nvidia for the provision of a set of Quadro GPU boards, and to Wolfram Research for {\it GridMathematica} provision to the King's College London Centre for Grid Financial Computation. I am also grateful to Con Keating and Patrick Burns for making me aware of industry technology on random portfolios.

\end{document}